\documentclass[a4paper]{article}
\usepackage{amsmath}
\usepackage{amssymb}
\usepackage{amsfonts}
\usepackage{amsthm}

\theoremstyle{plain}

\newtheorem*{thm*}{Theorem}

\theoremstyle{definition}

\usepackage[top=30truemm,bottom=30truemm,left=25truemm,right=25truemm]{geometry}

\usepackage{enumitem}
\usepackage{dcolumn}
\usepackage{longtable}

\usepackage{graphicx, color}

\begin{document}

\title{Empirical analysis in limit order book modeling for Nikkei 225 Stocks with Cox-type intensities
\footnote{
This work was in part supported by 
Japan Science and Technology Agency CREST JPMJCR2115; 
Japan Society for the Promotion of Science Grants-in-Aid for Scientific Research 
No. 17H01702 (Scientific Research);  
and by a Cooperative Research Program of the Institute of Statistical Mathematics. 
}
}
\author{Shunya Chomei}
\date{\today}
\maketitle

\noindent{\it Summary}
In this paper, we build on the analysis of Muni Toke and Yoshida \cite{muni2020analyzing} and conduct several empirical studies using high-frequency financial data.
Muni Toke and Yoshida \cite{muni2020analyzing} showed the consistency and asymptotic behavior of the Cox-type model estimators for relative intensities of orders in the limit order book, and then by using high-frequency trading data for $36$ stocks traded on the Paris Stock Exchange, they carry out model selection and trading sign prediction.
In this study, we add new covariates and carry out model selection and trading sign prediction using high-frequency trading data for $222$ stocks traded on the Tokyo Stock Exchange.
We not only show that the Cox-type model performs well in the Japanese market as well as in the Euronext Paris market, but also present the key factors for more accurate estimation.
We also suggest how often the covariates should be calibrated.

\section{Introduction}

The limit order book is a data set that records the time, volume, price, and other information of orders to buy or sell stocks traded in the real financial markets.
There are three types of orders: limit orders, market orders, and cancellations.
Limit orders are methods of placing orders by specifying buy or sell prices, while market orders are methods to buy or sell at the best quote without specifying prices.
Limit orders are removed from the book when they are canceled or matched and executed with market orders.
Sell (resp. Buy) limit orders are displayed on the ask (resp. bid) side, and buy (resp. sell) market orders are matched against the current best ask (resp. bid) orders.

Limit order book modeling is underway for applications to the development of high-frequency trading algorithms.
Chakraborti et al. \cite{chakraborti2011econophysics}, and Eisler et al. \cite{eisler2012price} performed mathematical modeling of limit and market orders to reveal the statistical properties of financial time series.
In the modeling, Hawks processes are used in Bacry et al. \cite{bacry2012non}, Muni Toke and Pomponio \cite{toke2012modelling}, Lallouache and Challet \cite{lallouache2016limits}, and Lu and Abergel \cite{lu2018high}.
Furthermore, Muni Toke and Yoshida \cite{muni2017modelling} and Morariu-Patrichi and Pakkanen \cite{morariu2022state} added state-dependent terms to the Hawkes process.
Also, Rambaldi et al. \cite{rambaldi2017role} used a marked multivariate Hawkes process.
Muni Toke and Yoshida \cite{muni2020analyzing} investigated the impact of financial variables (order book status and other trading signals) on the process of market orders.
They consider point processes of bid and ask market orders and assume that each intensity can be written as a Cox-type intensity, which is the product of a baseline intensity representing global market activity, etc., and a factor that depends on a given covariate.
By dealing with the ratio of the intensities, they remove the baseline intensity from the estimation procedure and estimate the relative influences of the covariates on the trading sign decision.
The factors that have a significant impact on the dynamics of the limit order book were revealed, and it was shown that it is possible to guess with a high degree of accuracy whether the next instant's market order is a bid or an ask for several stocks traded on the Paris Stock Exchange.

In this paper, we show that our algorithm works for multiple stocks traded on the Tokyo Stock Exchange.
Also, we show that adding the $n$-th imbalance (the fraction of stocks available for trading at the $n$-th best quote for both bid-ask) and historical imbalances to the covariates increases the estimation accuracy.
Furthermore, Muni Toke and Yoshida \cite{muni2020analyzing} used $1$ day of historical data as training data to estimate parameters and predict the next day's trading signs, but in this study, we compare the estimation accuracy of the same model for multiple numbers of days.
From these results, we show that the number of days of data used for estimation makes a difference in accuracy, suggesting the need for analysis of the look-back period.

\section{Description of the model and estimation procedure} \label{0501062323}

Let $\mathbb{I} = \{ \mathit{MA}, \mathit{MB} \}$ and $\mathbb{J} = \{1, ..., {\bar{j}}\}$ with $\bar{j} \in \mathbb{N}$.
Let $(N^{\mathit{MA}}_t)_{t \geq 0}$ and $(N^{\mathit{MB}}_t)_{t \geq 0}$ be point processes representing the number of ask and bid market orders, respectively, placed up to time $t$.
For $i \in \mathbb{I}$, an intensity $\lambda^i(t)$ of $N^i$ is written as

\begin{eqnarray}
\lambda^i(t, \vartheta) = \lambda_0(t) \exp \left( \sum_{j \in \mathbb{J}} \vartheta^i_j X_j(t) \right),
\end{eqnarray}
where $\lambda_0(t)$ is an unobservable common random baseline intensity, $(X_j(t))_{t \geq 0}$ is a $j$-th observable covariate process, and $\vartheta = (\vartheta^i_j)_{i \in \mathbb{I}, j \in \mathbb {J}}$ is a parameter.
Let $\mathbb{X}(t) = (X_j(t))_{j \in \mathbb{J}}$.

We are interested in the relative intensities, that is, the reference,

\begin{eqnarray}
\theta_j = \vartheta^{\mathit{MA}}_j - \vartheta^{\mathit{MB}}_j \qquad (j \in \mathbb{J}),
\end{eqnarray}
,not in the values of the coefficients $\vartheta^i_j$.
Since we are only interested in the relative intensities, we consider intensity ratios instead of the standard intensities defined by (1).

\begin{eqnarray}
r^i(t, \theta) &=& \frac{\lambda^i(t, \vartheta)}{\displaystyle \sum_{i' \in \mathbb{I}} \lambda^{i'}(t, \vartheta)}
= \frac{\exp \left(\displaystyle \sum_{j \in \mathbb{J}} \vartheta^i_j X_j(t) \right)}{\displaystyle \sum_{i' \in \mathbb{I}} \exp \left(\displaystyle \sum_{j \in \mathbb{J}} \vartheta^{i'}_j X_j(t) \right)} \nonumber \\
&=& \left[ 1 + \exp \left(\displaystyle \sum_{j \in \mathbb{J}} (\vartheta^{i'}_j - \vartheta^{i}_j) X_j(t) \right) \right]^{-1} \qquad i' \in \mathbb{I} \setminus \{ i \} \nonumber \\
&=&
\begin{cases}
  \left[ 1 + \displaystyle \exp \left(\displaystyle \sum_{j \in \mathbb{J}} \theta_j X_j(t) \right) \right]^{-1} & \text{if $i = \mathit{MB}$,} \\
  \left[ 1 + \displaystyle \exp \left(\displaystyle \sum_{j \in \mathbb{J}} -\theta_j X_j(t) \right) \right]^{-1} & \text{if $i = \mathit{MA}$.}
\end{cases}
\end{eqnarray}

Note that $r^{\mathit{MA}}(t, \theta) + r^{\mathit{MB}}(t, \theta) = 1$.
Let a bounded closed domain $\Theta \subset \mathbb{R}^{\bar{j}}$ be a parameter space of $\theta$.

We consider an observation sequence of intraday data.
In general, the observations are non-ergodic.
We take an intervals $I^{(k)} = [O_k, C_k], \, k = 1, \dots , T$ of the same length such that $0 \leq O_1 < C_1 \leq O_2 < C_2 \leq \dots $.
Now let $((N^i_t)_{i \in \mathbb{I}}, \, \mathbb{X}(t))_{t \in I^{(k)}}$, with $k = 1, \dots , T$ be the observations.

\begin{eqnarray}
\mathbb{H}_T(\theta) = \sum_{k=1}^T \sum_{i \in \mathbb{I}} \int_{I^{(k)}} \log r^i(t, \theta) dN^i_t.
\end{eqnarray}
We consider the quasi-log likelihood (log partial likelihood) to estimate the parameter $\theta$ defined by (2).
A quasi-maximum likelihood estimator (QMLE) is a measurable map $\hat{\theta}^M_T : \Omega \longrightarrow \Theta$ satisfying

\begin{eqnarray}
\mathbb{H}_T(\hat{\theta}^M_T ) = \max_{\theta \in \Theta} \mathbb{H}_T(\theta), \qquad \omega \in \Omega.
\end{eqnarray}
Let $\mathcal{X}_k = (\lambda_0(t), \mathbb{X}(t))_{t \in I^{(k)}}, \, \mathcal{G}_k = \sigma [\mathcal{X}_1, \dots, \mathcal{X}_k], \, \mathcal{H}_k = \sigma [\mathcal{X}_k, \mathcal{X}_{k+1}, \dots]$, and\\
$\displaystyle \alpha^{\mathcal{X}}(h) = \displaystyle \sup_{k \in \mathbb{N}} \displaystyle \sup_{A \in \mathcal{G}_k, \, B \in \mathcal{H}_{k+h}} |P[A \cap B] - P[A]P[B]|$.
We consider the following conditions:

\begin{description}
  \item[\textrm{[}C1\textrm{]}]
  $(\mathcal{X}_k)_{k \in \mathbb{N}}$ is identically distributed, $\displaystyle \sup_{t \in \mathbb{I}^{(1)}} || \lambda_0(t) ||_p < \infty$ and $\displaystyle \max_{j \in \mathbb{J}} \displaystyle \sup_{t \in \mathbb{I}^{(1)}} || \exp(p|X_j(t)|) ||_1 < \infty, \, \forall \, p > 1$.
  \item[\textrm{[}C2\textrm{]}]
  $\displaystyle \limsup_{h \rightarrow \infty} h^L \alpha^{\mathcal{X}}(h) < \infty, \, L > 0$.
\end{description}

Define the symmetric tensor $\Gamma$ by

\begin{eqnarray}
\Gamma [u^{\otimes2}] = E \left[ \int_{I^{(1)}} r^\mathit{MA}(t, \theta^*) \, r^\mathit{MB}(t, \theta^*) \, \mathbb{X}(t)^{\otimes2} [u^{\otimes2}] \Lambda(\lambda_0(t), \mathbb{X}(t)) \, dt \right], \qquad u \in \mathbb{R}^{\bar{j}},
\end{eqnarray}
where $\theta^* \in \Theta$ denotes the true value of $\theta$,
$\Lambda(w, x) = w \displaystyle \sum_{i \in \mathbb{I}} \exp \left(\displaystyle \sum_{j \in \mathbb{J}} x_j \vartheta^{*i}_j \right), \, x \in \mathbb{R}^{\bar{j}}$, and $\vartheta^*$ denotes the true value of $\vartheta$.
Here we further consider the following condition:

\begin{description}
  \item[\textrm{[}C3\textrm{]}]
  $\det \Gamma > 0$.
\end{description}

Let $\hat{u}^M_T = \sqrt{T} (\hat{\theta}^M_T - \theta^*)$.
Let $C_p(\mathbb{R}^{\bar{j}})$ be the space of continuous functions on $\mathbb{R}^{\bar{j}}$ of at most polynomial growth, and $\zeta$ a $\bar{j}$-dimensional standard Gaussian vector.
We obtain convergence of moments and asymptotic normality of the quasi-likelihood estimators.

\begin{thm*}
  Suppose that \textrm{[}C1\textrm{]}, [C2],and [C3] are satisfied.
  Then
  \begin{eqnarray}
  E \left[ f(\hat{u}^M_T) \right] \longrightarrow E \left[ f(\Gamma^{-1/2} \, \zeta) \right], \qquad \textrm{as} \quad T \longrightarrow \infty, \qquad f \in C_p(\mathbb{R}^{\bar{j}}).
  \end{eqnarray}
  (See the proof in Muni Toke and Yoshida \cite{muni2020analyzing}, Theorem 3.1.)
\end{thm*}

\section{Empirical study} \label{0501062324}

\subsection{Data}

We use individual stock tick data from the Nikkei NEEDS (Nikkei Economic Electronic Databank System) tick data file for $222$ stocks traded on the Tokyo Stock Exchange between March $2019$ and February $2020$.
More than $100$ million market orders and more than $1$ billion limit orders and cancellations in $222$ stocks are used for estimation.
The list of the stocks is given in Appendix.

\subsection{Description of covariates used}

We use the following covariates:

\begin{description}
  \item[Constant \, $ 1$]\mbox{}\\

  \item[The $n$-th imbalance \, $i_n(t)$]\mbox{}\\
  the ratio of the number of stocks available in the $n$-th quote for each ask-bid at time $t$.

  \begin{eqnarray}
  i_n(t) = \displaystyle \frac{q^B_n(t) - q^A_n(t)}{q^B_n(t) + q^A_n(t)}
  \end{eqnarray}
  where $q^A_n(t)$ and $q^B_n(t)$ are the number of stocks available in the $n$-th quote on the ask and bid sides at time t, respectively.
  The imbalance close to +1 (resp. -1) indicates that there is very little volume available on the ask (resp. bid) side, and that prices are likely to rise (resp. fall).

  \item[The imbalance of cumulative amount up to the $n$-th quote \, $\overline{i_n}(t)$]\mbox{}\\
  The imbalance at time $t$ is calculated using the cumulative amount of stocks available in quotes up to the $n$-th level for each ask-bid.
  ($i_1(t) = \overline{i_1}(t)$)

  \begin{eqnarray}
  \overline{i_n}(t) = \displaystyle \frac{\displaystyle \sum_{k=1}^n q^B_k(t) - \displaystyle \sum_{k=1}^n q^A_k(t)}{\displaystyle \sum_{k=1}^n q^B_k(t) + \displaystyle \sum_{k=1}^n q^A_k(t)}.
  \end{eqnarray}

  \item[Sign of the last trade \, $\epsilon(t)$]\mbox{}\\
  Let $\epsilon(t)$ be sign of the last market order at time $t$.
  ($\epsilon(t) = -1$ for an ask trade, $\epsilon(t) = +1$ for a bid trade)

  \item[Product of spread and last signs \, $\epsilon(t)s(t)$]\mbox{}\\
  The product of $\epsilon(t)$ and $s(t)$, where $s(t) = +1$ if the bid-ask spread is larger than its mean, and $-1$ if it is smaller.

  \item[The past n-th Imbalance \, $i_n(t^m)$]\mbox{}\\
  The n-th imbalance $i_n(t^m)$, where $t^m$ is the time when the market order was submitted $m$ times ago from the current time $t$.

  \item[The past imbalance of cumulative amount up to the $n$-th quote \, $\overline{i_n}(t^m)$]\mbox{}\\
  The imbalance of cumulative amount up to the $n$-th quote $\overline{i_n}(t^m)$, where $t^m$ is the time when the market order was submitted $m$ times ago from the current time $t$.

\end{description}

\subsection{Empirical result 1: Prediction}\label{0501062113}

In this section we will explain the potential use of the ratio model as a prediction tool.

After calculating QMLE using the previous day's LOB data, the ratio $r^i(t, \theta)$ of market order occurrence is calculated from the day's data.
We predict that the next market order will be on the ask side if the probability $r^{\mathit{MA}}(t, \theta)$ on the ask side is higher than 0.5; otherwise, we predict that it will be on the bid side.
We will compare the accuracy of the predicted trade sign by models, then also compare the accuracy when the signs alters by models.
Here only the "$l$ day" model uses the parameters obtained with $l$ days of LOB data to predict the next $l$ days.

The following covariates are used for each model:

\begin{longtable}[c]{|l|l||l|}
  \hline
  \multicolumn{2}{|l||}{Model} & Covariates\\
  \hline\hline
  imb $n$ & $n=1,2,\dots10$ & $1, \, i_1(t), \, i_2(t), \, \dots, \, i_n(t)$\\
  \hline
  imb $n$\_sum & $n=1,2,\dots10$ & $1, \, \overline{i_1}(t), \, \overline{i_2}(t), \, \dots, \, \overline{i_n}(t)$\\
  \hline
  imb $n$\_la $1$ & $n=1,2,\dots5$ & $1, \, i_1(t), \, i_2(t), \, \dots, \, i_n(t), \, i_1(t^1), \, i_2(t^1), \, \dots, \, i_n(t^1)$\\
  \hline
  imb $n$\_la $1$\_sum & $n=1,2,\dots5$ & $1, \, \overline{i_1}(t), \, \overline{i_2}(t), \, \dots, \, \overline{i_n}(t), \, \overline{i_1}(t^1), \, \overline{i_2}(t^1), \, \dots, \, \overline{i_n}(t^1)$\\
  \hline
  imb $n$\_e\_es & $n=1,2,\dots5$ & $1, \, i_1(t), \, i_2(t), \, \dots, \, i_n(t), \, \epsilon, \, \epsilon s$\\
  \hline
  imb $n$\_e\_es\_sum & $n=1,2,\dots5$ & $1, \, \overline{i_1}(t), \, \overline{i_2}(t), \, \dots, \, \overline{i_n}(t), \, \epsilon, \, \epsilon s$\\
  \hline
  imb $n$\_e\_es\_la $1$ & $n=1,2,\dots5$ & $1, \, i_1(t), \, i_2(t), \, \dots, \, i_n(t), \, i_1(t^1), \, i_2(t^1), \, \dots, \, i_n(t^1), \, \epsilon, \, \epsilon s$\\
  \hline
  imb $n$\_e\_es\_la $1$\_sum & $n=1,2,\dots5$ & $1, \, \overline{i_1}(t), \, \overline{i_2}(t), \, \dots, \, \overline{i_n}(t), \, \overline{i_1}(t^1), \, \overline{i_2}(t^1), \, \dots, \, \overline{i_n}(t^1), \, \epsilon, \, \epsilon s$\\
  \hline
  imb $n$\_e\_es\_la $m$ & $n=1,2$ & $1, \, i_1(t), \, \dots, \, i_n(t), \, i_1(t^1), \, \dots, \, i_n(t^1),$\\
  & $m=1,2,\dots5$ & $\hfill \dots, \, i_1(t^m), \, \dots, \, i_n(t^m), \, \epsilon, \, \epsilon s$\\
  \hline
  imb $n$\_e\_es\_la $1$\_ $l$day & $n=1,2$ & $1, \, i_1(t), \, \dots, \, i_n(t), \, i_1(t^1), \, \dots, \, i_n(t^1), \, \epsilon, \, \epsilon s$\\
  & $l=2,3,5,7,10,14,30,60$ & \hfill (Calibrate every $l$ days)\\
  \hline
\end{longtable}

We note that "imb $1$\_..." model and "imb $1$\_...\_sum" model are the same model.
The results for the accuracy for each model are shown in Figures 1 and 2.

\begin{figure}[htbp]
  \renewcommand{\figurename}{Figure }
  \begin{minipage}[htbp]{0.45\linewidth}
    \centering
    \includegraphics[keepaspectratio,scale=0.8,width=560pt,angle=90]{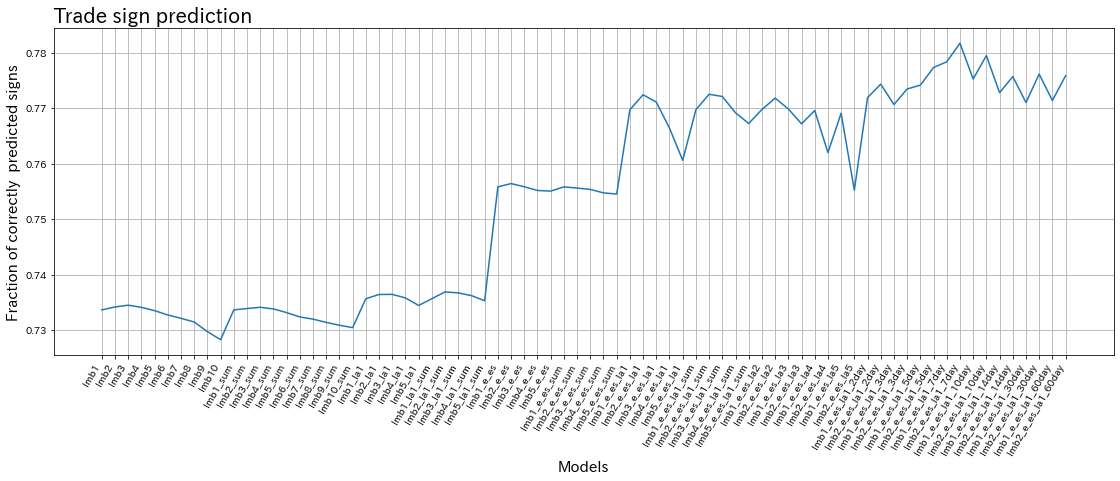}
    \caption{the accuracy}
  \end{minipage}
  \hspace{0.1\columnwidth}
  \begin{minipage}[htbp]{0.45\linewidth}
    \centering
    \includegraphics[keepaspectratio,scale=0.8,width=560pt,angle=90]{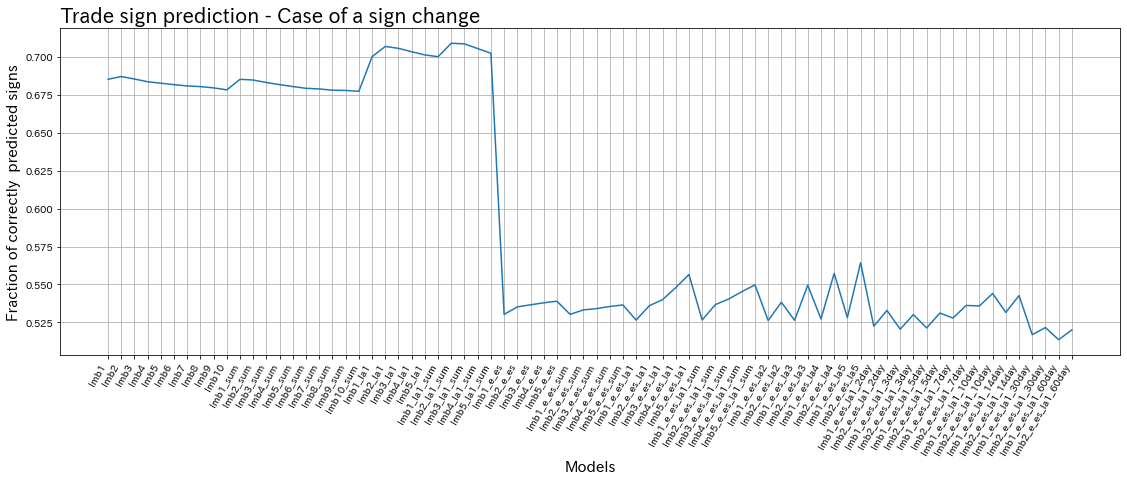}
    \caption{the accuracy when the signs alter}
  \end{minipage}
\end{figure}

The imbalance $i_n$ is a valid covariate and the accuracy for the “imb $n$” model is about $73\%$.
The accuracy increases slightly when the imbalance at prices that are $1$ - $2$ ticks away from the best quote is included in the covariates.
The method of calculating the imbalance is not much different in either case.
The imbalance $i_1^1$ when the $1$ previous market order was submitted, the last trade sign $\epsilon$, and the product of the spread and the last sign $\epsilon s$ are also valid, so the accuracy of “imb $n$\_la $1$” model, ” imb $n$\_e\_es” model, and “imb $n$\_e\_es\_la $1$” model is about $73.5\%$, about $75.5\%$, and about $77\%$, respectively.
On the other hand, including the imbalance at prices more than $3$ ticks away from the best quote or more than $2$ times prior in the covariates does not change or slightly reduces the accuracy.

The accuracy for the case of sign change for the imbalance-only model is about $68\%$ and for the model including past imbalances is more than $70\%$, but the accuracy drops significantly to $50\%-55\%$ when the last trade sign and spread are included.
This may be due to the fact that orders on the same side tend to be consecutive, and thus the prediction is likely to be dragged down by the previous trade sign if the last trade sign is included in the covariate.

Also, in the “imb $n$\_e\_es\_la $1$ \quad ($n=1,2$)” model, the accuracy can be improved from about $77\%$ to about $78\%$ by calibrating every $1-2$ weeks instead of daily.
This means that the more training data there are, the closer the estimates approach true value.
On the other hand, the accuracy decreased when the time period for parameter estimation was increased, suggesting that the true values of the parameters fluctuated over time and structural changes occurred.

\subsection{Empirical result 2: Model Selection}

This section describes the model selection using the information criterion.
A mathematical validation of the use of an information criterion can be found in Muni Toke and Yoshida \cite{muni2020analyzing}, Chapter 4.
We confirm that models with lower values of the information criteria are also more accurate in predicting trading signs.

We use the following $3$ information criterion:

\begin{description}[labelwidth=20em]
  \item[the quasi-AIC (QAIC)] $-2 \mathbb{H}_T(\hat{\theta}^M_T) + 2 d$,
  \item[the quasi-consistent AIC (QCAIC)] $-2 \mathbb{H}_T(\hat{\theta}^M_T) + (\log T + 1) d$,
  \item[the quasi-BIC (QBIC)] $-2 \mathbb{H}_T(\hat{\theta}^M_T) + (\log T )d$,
\end{description}
where $d$ is a dimension on the parameter space of $\theta$.

\begin{figure}[htbp]
  \renewcommand{\figurename}{Figure }
  \begin{minipage}[htbp]{1\linewidth}
    \centering
    \includegraphics[keepaspectratio,scale=0.8,width=650pt,angle=90]{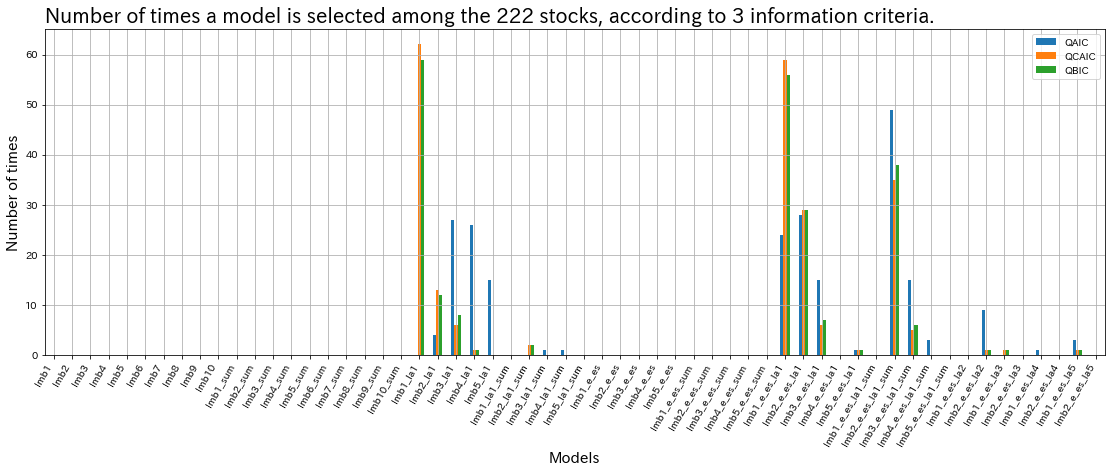}
    \caption{Number of times a model is selected among the 222 stocks, according to 3 information criteria.}\label{0501062123}
  \end{minipage}
\end{figure}

The number of times a model was selected from among the models except "imb $n$\_e\_es\_la $1$\_ $l$day" model for $222$ stocks.
Results are shown in Figure \ref{0501062123}.
The reason why the I”mb $n$\_e\_es\_la $1$\_ $l$day” model was not used is that $T$ takes different values for different periods used for parameter estimation, making it impossible to simply compare the values of the information criterion.

The models that are selected the most often are “imb $1$\_la $1$” model, “imb $1$\_e\_es\_la $1$” model, “imb $2$\_e\_es\_la $1$\_sum” model, and “imb $2$\_e\_es\_la $1$” model.
It can be seen that many models with a small number of parameters are selected among the models with high accuracy in chapter \ref{0501062113}.
“Imb $1$\_la $1$” model can be evaluated to some extent as a model with a reasonably high accuracy despite its small number of parameters.
This model is selected by QCAIC and QBIC, indicating that QCAIC and QBIC tend to select models with fewer parameters.
On the other hand, "imb $1$\_e\_es\_la $1$" model, "imb $2$\_e\_es\_la $1$\_sum" model, and "imb $2$\_e\_es\_la $1$" model have a high accuracy of $77\%$, and we can say that the model selection is working well.
Models with large parameters such as "imb $2$\_e\_es\_la $1$\_sum" model and "imb $2$\_e\_es\_la $1$" model are selected relatively often by QAIC.

\newpage

\section{Conclusion}

We conducted an empirical study using a model based on ratios of Cox-type intensities sharing a common random baseline intensity proposed in Muni Toke and Yoshida \cite{muni2020analyzing}
It is shown that the ratio model can predict the next trading sign with good accuracy using a recent order book, even in the Japanese market.
Furthermore, by using the past imbalance as a new covariate, we can improve the accuracy of the prediction.
Future research includes discovering additional covariates to identify important trading signals, finding a method to estimate the appropriate number of look-back days, and constructing an algorithm for high-frequency trading that predicts actual price fluctuations by taking into account the aggressiveness of market orders.

\bibliographystyle{spmpsci}      
\bibliography{bibtex220423-220512}   

\section{Appendix}

We use the following stocks in this study:

\newpage

\footnotesize

\begin{longtable}[c]{|l||l|l|}
  \caption{List of stocks}\\
  \hline
  Code&Issue name&Industry\\
  \hline\hline
  1332&Nippon Suisan Kaisha,Ltd.&Fishery, Agriculture and Forestry\\
  1333&Maruha Nichiro Corporation&Fishery, Agriculture and Forestry\\
  1605&INPEX CORPORATION&Mining\\
  1721&COMSYS Holdings Corporation&Construction\\
  1801&TAISEI CORPORATION&Construction\\
  1802&OBAYASHI CORPORATION&Construction\\
  1803&SHIMIZU CORPORATION&Construction\\
  1808&HASEKO Corporation&Construction\\
  1812&KAJIMA CORPORATION&Construction\\
  1925&DAIWA HOUSE INDUSTRY CO.,LTD.&Construction\\
  1928&Sekisui House,Ltd.&Construction\\
  1963&JGC HOLDINGS CORPORATION&Construction\\
  2002&NISSHIN SEIFUN GROUP INC.&Foods\\
  2269&Meiji Holdings Co.,Ltd.&Foods\\
  2282&NH Foods Ltd.&Foods\\
  2432&DeNA Co.,Ltd.&Services\\
  2501&SAPPORO HOLDINGS LIMITED&Foods\\
  2502&Asahi Group Holdings,Ltd.&Foods\\
  2503&Kirin Holdings Company,Limited&Foods\\
  2531&TAKARA HOLDINGS INC.&Foods\\
  2768&Sojitz Corporation&Wholesale Trade\\
  2801&KIKKOMAN CORPORATION&Foods\\
  2802&Ajinomoto Co.,Inc.&Foods\\
  2871&NICHIREI CORPORATION&Foods\\
  2914&JAPAN TOBACCO INC.&Foods\\
  3086&J.FRONT RETAILING Co.,Ltd.&Retail Trade\\
  3099&Isetan Mitsukoshi Holdings Ltd.&Retail Trade\\
  3101&TOYOBO CO.,LTD.&Textiles and Apparels\\
  3103&UNITIKA LTD.&Textiles and Apparels\\
  3105&Nisshinbo Holdings Inc.&Electric Appliances\\
  3289&Tokyu Fudosan Holdings Corporation&Real Estate\\
  3382&Seven and I Holdings Co.,Ltd.&Retail Trade\\
  3401&TEIJIN LIMITED&Textiles and Apparels\\
  3402&TORAY INDUSTRIES,INC.&Textiles and Apparels\\
  3405&KURARAY CO.,LTD.&Chemicals\\
  3407&ASAHI KASEI CORPORATION&Chemicals\\
  3436&SUMCO CORPORATION&Metal Products\\
  3861&Oji Holdings Corporation&Pulp and Paper\\
  3863&Nippon Paper Industries Co.,Ltd.&Pulp and Paper\\
  4004&Showa Denko K.K.&Chemicals\\
  4005&SUMITOMO CHEMICAL COMPANY,LIMITED&Chemicals\\
  4021&Nissan Chemical Corporation&Chemicals\\
  4042&TOSOH CORPORATION&Chemicals\\
  4043&Tokuyama Corporation&Chemicals\\
  4061&Denka Company Limited&Chemicals\\
  4063&Shin-Etsu Chemical Co.,Ltd.&Chemicals\\
  4151&Kyowa Kirin Co.,Ltd.&Pharmaceutical\\
  4183&Mitsui Chemicals,Inc.&Chemicals\\
  4188&Mitsubishi Chemical Holdings Corporation&Chemicals\\
  4208&UBE Corporation&Chemicals\\
  4272&NIPPON KAYAKU CO.,LTD.&Chemicals\\
  4324&DENTSU GROUP INC.&Services\\
  4452&Kao Corporation&Chemicals\\
  4502&Takeda Pharmaceutical Company Limited&Pharmaceutical\\
  4503&Astellas Pharma Inc.&Pharmaceutical\\
  4506&Sumitomo Pharma Co.,Ltd.&Pharmaceutical\\
  4507&Shionogi and Co.,Ltd.&Pharmaceutical\\
  4519&CHUGAI PHARMACEUTICAL CO.,LTD.&Pharmaceutical\\
  4523&Eisai Co.,Ltd.&Pharmaceutical\\
  4543&TERUMO CORPORATION&Precision Instruments\\
  4568&DAIICHI SANKYO COMPANY,LIMITED&Pharmaceutical\\
  4578&Otsuka Holdings Co.,Ltd.&Pharmaceutical\\
  4689&Z Holdings Corporation&Services\\
  4704&Trend Micro Incorporated&Services\\
  4755&Rakuten Group,Inc.&Services\\
  4901&FUJIFILM Holdings Corporation&Chemicals\\
  4902&KONICA MINOLTA,INC.&Precision Instruments\\
  4911&Shiseido Company,Limited&Chemicals\\
  5020&ENEOS Holdings,Inc.&Oil and Coal Products\\
  5101&The Yokohama Rubber Company,Limited&Rubber Products\\
  5108&BRIDGESTONE CORPORATION&Rubber Products\\
  5201&AGC Inc.&Glass and Ceramics Products\\
  5202&Nippon Sheet Glass Company,Limited&Glass and Ceramics Products\\
  5214&Nippon Electric Glass Co.,Ltd.&Glass and Ceramics Products\\
  5232&Sumitomo Osaka Cement Co.,Ltd.&Glass and Ceramics Products\\
  5233&TAIHEIYO CEMENT CORPORATION&Glass and Ceramics Products\\
  5301&TOKAI CARBON CO.,LTD.&Glass and Ceramics Products\\
  5332&TOTO LTD.&Glass and Ceramics Products\\
  5333&NGK INSULATORS,LTD.&Glass and Ceramics Products\\
  5401&NIPPON STEEL CORPORATION&Iron and Steel\\
  5406&Kobe Steel,Ltd.&Iron and Steel\\
  5411&JFE Holdings,Inc.&Iron and Steel\\
  5541&PACIFIC METALS CO.,LTD.&Iron and Steel\\
  5631&The Japan Steel Works,Ltd.&Machinery\\
  5703&Nippon Light Metal Holdings Company,Ltd.&Metal Products\\
  5706&Nippon Light Metal Holdings Company,Ltd.&Metal Products\\
  5707&Toho Zinc CO.,Ltd.&Metal Products\\
  5711&Mitsubishi Materials Corporation&Metal Products\\
  5713&Sumitomo Metal Mining Co.,Ltd.&Metal Products\\
  5714&DOWA HOLDINGS CO.,LTD.&Metal Products\\
  5715&FURUKAWA CO.,LTD.&Metal Products\\
  5801&Furukawa Electric Co.,Ltd.&Metal Products\\
  5802&Sumitomo Electric Industries,Ltd.&Metal Products\\
  5803&Fujikura Ltd.&Metal Products\\
  5901&Toyo Seikan Group Holdings,Ltd.&Metal Products\\
  6098&Recruit Holdings Co.,Ltd.&Services\\
  6103&OKUMA Corporation&Machinery\\
  6113&AMADA CO.,LTD.&Machinery\\
  6178&JAPAN POST HOLDINGS Co.,Ltd.&Services\\
  6301&KOMATSU LTD.&Machinery\\
  6302&SUMITOMO HEAVY INDUSTRIES,LTD.&Machinery\\
  6305&Hitachi Construction Machinery Co.,Ltd.&Machinery\\
  6326&KUBOTA CORPORATION&Machinery\\
  6361&EBARA CORPORATION&Machinery\\
  6366&Chiyoda Corporation&Machinery\\
  6367&DAIKIN INDUSTRIES,LTD.&Machinery\\
  6471&NSK Ltd.&Machinery\\
  6472&NTN CORPORATION&Machinery\\
  6473&JTEKT Corporation&Machinery\\
  6479&MINEBEA MITSUMI Inc.&Electric Appliances\\
  6501&Hitachi,Ltd.&Electric Appliances\\
  6503&Mitsubishi Electric Corporation&Electric Appliances\\
  6504&FUJI ELECTRIC CO.,LTD.&Electric Appliances\\
  6506&YASKAWA Electric Corporation&Electric Appliances\\
  6674&GS Yuasa Corporation&Electric Appliances\\
  6701&NEC Corporation&Electric Appliances\\
  6702&FUJITSU LIMITED&Electric Appliances\\
  6703&Oki Electric Industry Company,Limited&Electric Appliances\\
  6724&SEIKO EPSON CORPORATION&Electric Appliances\\
  6752&Panasonic Holdings Corporation&Electric Appliances\\
  6758&SONY GROUP CORPORATION&Electric Appliances\\
  6762&TDK Corporation&Electric Appliances\\
  6770&ALPS ALPINE CO.,LTD.&Electric Appliances\\
  6841&YOKOGAWA ELECTRIC CORPORATION&Electric Appliances\\
  6857&ADVANTEST CORPORATION&Electric Appliances\\
  6902&DENSO CORPORATION&Electric Appliances\\
  6952&CASIO COMPUTER CO.,LTD.&Electric Appliances\\
  6954&FANUC CORPORATION&Electric Appliances\\
  6971&KYOCERA CORPORATION&Electric Appliances\\
  6976&TAIYO YUDEN CO.,LTD.&Electric Appliances\\
  6988&NITTO DENKO CORPORATION&Chemicals\\
  7003&Mitsui EandS Holdings Co.,Ltd.&Machinery\\
  7004&Hitachi Zosen Corporation&Machinery\\
  7011&Mitsubishi Heavy Industries,Ltd.&Machinery\\
  7012&Kawasaki Heavy Industries,Ltd.&Transportation Equipment\\
  7013&IHI Corporation&Machinery\\
  7186&Concordia Financial Group,Ltd.&Banks\\
  7201&NISSAN MOTOR CO.,LTD.&Transportation Equipment\\
  7202&ISUZU MOTORS LIMITED&Transportation Equipment\\
  7203&TOYOTA MOTOR CORPORATION&Transportation Equipment\\
  7205&HINO MOTORS,LTD.&Transportation Equipment\\
  7211&MITSUBISHI MOTORS CORPORATION&Transportation Equipment\\
  7261&Mazda Motor Corporation&Transportation Equipment\\
  7267&HONDA MOTOR CO.,LTD.&Transportation Equipment\\
  7269&SUZUKI MOTOR CORPORATION&Transportation Equipment\\
  7270&SUBARU CORPORATION&Transportation Equipment\\
  7272&Yamaha Motor Co.,Ltd.&Transportation Equipment\\
  7731&NIKON CORPORATION&Precision Instruments\\
  7733&OLYMPUS CORPORATION&Precision Instruments\\
  7735&SCREEN Holdings Co.,Ltd.&Electric Appliances\\
  7751&CANON INC.&Electric Appliances\\
  7752&RICOH COMPANY,LTD.&Electric Appliances\\
  7762&Citizen Watch Co.,Ltd.&Precision Instruments\\
  7911&TOPPAN INC.&Other Products\\
  7912&Dai Nippon Printing Co.,Ltd.&Other Products\\
  7951&YAMAHA CORPORATION&Other Products\\
  8001&ITOCHU Corporation&Wholesale Trade\\
  8002&Marubeni Corporation&Wholesale Trade\\
  8015&TOYOTA TSUSHO CORPORATION&Wholesale Trade\\
  8028&FamilyMart Co.,Ltd.&Retail Trade\\
  8031&MITSUI and CO.,LTD.&Wholesale Trade\\
  8035&Tokyo Electron Limited&Electric Appliances\\
  8053&SUMITOMO CORPORATION&Wholesale Trade\\
  8058&Mitsubishi Corporation&Wholesale Trade\\
  8233&Takashimaya Company,Limited&Retail Trade\\
  8252&MARUI GROUP CO.,LTD.&Retail Trade\\
  8253&Credit Saison Co.,Ltd.&Other Financing Business\\
  8267&AEON CO.,LTD.&Retail Trade\\
  8303&Shinsei Bank,Limited&Banks\\
  8304&Aozora Bank,Ltd.&Banks\\
  8306&Mitsubishi UFJ Financial Group,Inc.Resona Holdings, Inc.&Banks\\
  8308&Resona Holdings, Inc.&Banks\\
  8309&Sumitomo Mitsui Trust Holdings,Inc.&Banks\\
  8316&Sumitomo Mitsui Financial Group,Inc.&Banks\\
  8331&The Chiba Bank,Ltd.&Banks\\
  8354&Fukuoka Financial Group,Inc.&Banks\\
  8355&THE SHIZUOKA BANK,LTD.&Banks\\
  8411&Mizuho Financial Group,Inc.&Banks\\
  8601&Daiwa Securities Group Inc.&Securities and Commodity Futures\\
  8604&Nomura Holdings, Inc.&Securities and Commodity Futures\\
  8628&MATSUI SECURITIES CO.,LTD.&Securities and Commodity Futures\\
  8630&Sompo Holdings,Inc.&Insurance\\
  8725&MSandAD Insurance Group Holdings,Inc.&Insurance\\
  8729&Sony Financial Holdings Inc.&Insurance\\
  8750&Dai-ichi Life Holdings,Inc.&Insurance\\
  8766&Tokio Marine Holdings,Inc.&Insurance\\
  8795&TandD Holdings, Inc.&Insurance\\
  8801&Mitsui Fudosan Co.,Ltd.&Real Estate\\
  8802&Mitsubishi Estate Company,Limited&Real Estate\\
  8804&Tokyo Tatemono Co.,Ltd.&Real Estate\\
  8830&Sumitomo Realty and Development Co.,Ltd.&Real Estate\\
  9001&TOBU RAILWAY CO.,LTD.&Land Transportation\\
  9005&TOKYU CORPORATION&Land Transportation\\
  9007&Odakyu Electric Railway Co.,Ltd.&Land Transportation\\
  9008&Keio Corporation&Land Transportation\\
  9009&Keisei Electric Railway Co.,Ltd.&Land Transportation\\
  9020&East Japan Railway Company&Land Transportation\\
  9021&West Japan Railway Company&Land Transportation\\
  9022&Central Japan Railway Company&Land Transportation\\
  9062&Nippon Express Holdings,Inc.&Land Transportation\\
  9064&YAMATO HOLDINGS CO.,LTD.&Land Transportation\\
  9101&Nippon Yusen Kabushiki Kaisha&Marine Transportation\\
  9104&Mitsui O.S.K.Lines,Ltd.&Marine Transportation\\
  9107&Kawasaki Kisen Kaisha,Ltd.&Marine Transportation\\
  9202&ANA HOLDINGS INC.&Air Transportation\\
  9301&Mitsubishi Logistics Corporation&Warehousing and Harbor Transportation Services\\
  9412&SKY Perfect JSAT Holdings Inc.&Information and Communication\\
  9432&NIPPON TELEGRAPH AND TELEPHONE CORPORATION&Information and Communication\\
  9433&KDDI CORPORATION&Information and Communication\\
  9437&NTT DOCOMO&Information and Communication\\
  9501&Tokyo Electric Power Company Holdings,Incorporated&Electric Power and Gas\\
  9502&Chubu Electric Power Company,Incorporated&Electric Power and Gas\\
  9503&The Kansai Electric Power Company,Incorporated&Electric Power and Gas\\
  9531&TOKYO GAS CO.,LTD.&Electric Power and Gas\\
  9532&OSAKA GAS CO.,LTD.&Electric Power and Gas\\
  9602&TOHO CO.,LTD&Information and Communication\\
  9613&NTT DATA CORPORATION&Information and Communication\\
  9681&TOKYO DOME Services&Services\\
  9735&SECOM CO.,LTD.&Services\\
  9766&KONAMI HOLDINGS CORPORATION&Information and Communication\\
  9983&FAST RETAILING CO.,LTD.&Retail Trade\\
  9984&SoftBank Group Corp.&Information and Communication\\
  \hline
\end{longtable}

\end{document}